# THE AI ASSESSMENT SCALE REVISITED: A FRAMEWORK FOR EDUCATIONAL ASSESSMENT

Mike Perkins [1*], Jasper Roe [2], Leon Furze [3]

[1] British University Vietnam, Vietnam
[2] James Cook University Singapore, Singapore
[3] Deakin University, Australia

[*] Corresponding Author: Mike.p@buv.edu.vn

December 2024

| | | |
|---|---|---|
| 1 | NO AI | The assessment is completed entirely without AI assistance in a controlled environment, ensuring that students rely solely on their existing knowledge, understanding, and skills. You must not use AI at any point during the assessment. You must demonstrate your core skills and knowledge. |
| 2 | AI PLANNING | AI may be used for pre-task activities such as brainstorming, outlining and initial research. This level focuses on the effective use of AI for planning, synthesis, and ideation, but assessments should emphasise the ability to develop and refine these ideas independently. You may use AI for planning, idea development, and research. Your final submission should show how you have developed and refined these ideas. |
| 3 | AI COLLABORATION | AI may be used to help complete the task, including idea generation, drafting, feedback, and refinement. Students should critically evaluate and modify the AI suggested outputs, demonstrating their understanding. You may use AI to assist with specific tasks such as drafting text, refining and evaluating your work. You must critically evaluate and modify any AI-generated content you use. |
| 4 | FULL AI | AI may be used to complete any elements of the task, with students directing AI to achieve the assessment goals. Assessments at this level may also require engagement with AI to achieve goals and solve problems. You may use AI extensively throughout your work either as you wish, or as specifically directed in your assessment. Focus on directing AI to achieve your goals while demonstrating your critical thinking. |
| 5 | AI EXPLORATION | AI is used creatively to enhance problem-solving, generate novel insights, or develop innovative solutions to solve problems. Students and educators co-design assessments to explore unique AI applications within the field of study. You should use AI creatively to solve the task, potentially co-designing new approaches with your instructor. |

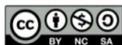

# Abstract

Recent developments in Generative Artificial Intelligence (GenAI) have created significant uncertainty in education, particularly in terms of assessment practices. Against this backdrop, we present an updated version of the AI Assessment Scale (AIAS), a framework with two fundamental purposes: to facilitate open dialogue between educators and students about appropriate GenAI use and to support educators in redesigning assessments in an era of expanding AI capabilities.

Grounded in social constructivist principles and designed with assessment validity in mind, the AIAS provides a structured yet flexible approach that can be adapted across different educational contexts. Building on implementation feedback from global adoption across both the K-12 and higher education contexts, this revision represents a significant change from the original AIAS. Among these changes is a new visual guide that moves beyond the original traffic light system and utilises a neutral colour palette that avoids implied hierarchies between the levels. The scale maintains five distinct levels of GenAI integration in assessment, from "No AI" to "AI Exploration", but has been refined to better reflect rapidly advancing technological capabilities and emerging pedagogical needs.

This paper presents the theoretical foundations of the revised framework, provides detailed implementation guidance through practical vignettes, and discusses its limitations and future directions. As GenAI capabilities continue to expand, particularly in multimodal content generation, the AIAS offers a starting point for reimagining assessment design in an era of disruptive technologies.

# Introduction

When we began working on the original AI Assessment Scale (AIAS) (Perkins, Furze, et al., 2024), the education sector faced a high degree of uncertainty. At that time, a new breed of chatbot powered by Large Language Models (LLMs) had emerged, with capabilities that exceeded those of the prior generation of technological support tools in academic work, such as Automated Paraphrasing Tools (APTs) (Prentice & Kinden, 2018; Roe & Perkins, 2022) or Digital Writing Assistants (DWA) (Roe et al., 2023). The initial response to these technologies included attempts to ban and detect their use in educational environments. However, as the technology grew in ubiquity, the discourse shifted from prohibition to cautious integration, based on the idea that educators and students must learn to work with what we now call Generative AI (GenAI) tools rather than against them. In 2023, calls for guidelines on how students should use GenAI ethically began to emerge (Cotton et al., 2023; Crawford et al., 2023; Perkins, 2023). Therefore, we decided to develop a framework to structure the cautious integration of AI tools into educational assessment, which we labelled the AI Assessment Scale (AIAS). Our goal at that time was to offer a way for educators, including ourselves, to have open dialogue with students about the values of educational integrity, and to prioritise a transparent approach to how GenAI could be used in assessment.

The AIAS has filled an urgent need for a flexible, transparent, and practical tool to help educators and students deal with GenAI. To date, the AIAS has been used in over a dozen countries and has been translated into at least 12 languages that the authors are aware of (Furze, 2024). We presented the AIAS during the UNESCO Digital Learning Week and introduced it to educators globally through numerous invited talks, workshops, and conferences. This collective experience has been incredibly valuable, and we have been able to have many rich and enlightening discussions about how the AIAS has been adapted, remixed, and developed in multiple contexts. At the same time, the creative applications of the AIAS, along with well-founded criticisms of our initial work, have helped to guide our thinking in refining and future-oriented frameworks.

In the remainder of this article, we discuss the current context of GenAI and education, and the staggering changes that have occurred in the last two years. We then reflect on the different applications of the AIAS thus far, and in doing so, address the feedback and critiques that the framework has received since its inception. This forms the basis of the third section, in which we introduce our second iteration of the AIAS, explain how and why the decisions were made to change and augment the original scale, and introduce a further level. Finally, we consider the future of assessment and attempt to justify why the AIAS may continue to be a useful tool in guiding assessment strategies and reform in the coming years.

## GenAI, Education and Assessment: The Story So Far

In 2022, the release of ChatGPT (with GPT-3.5) was met with concerns of academic integrity, with a focus on written assessments, such as essays (Cotton et al., 2023; Perkins, 2023; Roe et al., 2023). Since the initial release of ChatGPT, this technology has become much more capable across an ever-broadening range of mathematical and scientific reasoning tasks. The integration of GPT-4v vision capabilities (OpenAI, 2023), along with developments in image generation through open-source models such as Flux, the multimodal o1 model, and OpenAI's Sora video generation system, demonstrates that GenAI now impacts almost all disciplines and fields. Research has shown that GPT-4 can perform on par with human performance in a wide variety of multiple-choice question (MCQ) assessments, including high-stakes professional examinations in the fields of medicine, law, and science (Head & Willis, 2024; Newton et al., 2024). The newest iteration, ChatGPT-4o, scored 94% on the United Kingdom Medical Licencing Examination Applied Knowledge Test (Newton et al., 2024). GenAI models can now handle multimodal content effectively (Shahriar et al., 2024) and spend more time 'thinking' or 'reasoning', enabling them to achieve top-percentile scores in mathematical and coding competitions and challenges (OpenAI, 2024).

**The challenges of AI text detectors**

In response to the advent of GenAI tools and their ability to destabilise many forms of educational assessment, bans on the use of GenAI were quickly followed by AI detection tools brought to the market by several private corporations. Although we recognise that AI detectors may have a place in educational assessment, these should be primarily for providing feedback to students or having a conversation with students – not as an adversarial tool for catching students in the act of so-called cheating; an approach which Moro (2020) bluntly refers to as "cop shit". Another rationale for this stance is that AI detectors are unreliable (Elkhatat et al., 2023; Liang et al., 2023; Perkins et al., 2023; Perkins, Roe, et al., 2024; Sadasivan et al., 2023; Weber-Wulff et al., 2023), leading to unjustified and potentially harrowing consequences for students that significantly impact their academic and personal lives (Gorichanaz, 2023; Roe, Perkins, & Ruelle, 2024).

The development of adversarial discourse is also common in university policies, which frame the use of AI as misconduct (Luo (Jess), 2024a), and among instructors, who may now view students' produced texts more critically and with suspicion (Farazouli et al., 2023), leading to students using GenAI tools in an ethical and allowed way to hide their usage (Gonsalves, 2024). As such, the AIAS is intended to mitigate the impact of these adversarial relationships and their impact on student-teacher trust. A lack of transparency in how educators use or evaluate GenAI in the grading process impacts students' trust in the educational process (Luo (Jess), 2024b). However, there seems to be an expanding awareness of the fruitlessness of this approach, as it seems likely that hybrid writing or co-creation may soon become a new norm, and trying to distinguish between AI and non-AI enhanced writing could become a meaningless task (Eaton, 2023). Ultimately, AI detection is impossible and inefficient (Mao et al., 2024).

**A growing consensus**

There is now a consensus that for better or worse, we must learn to live with the reality that GenAI will be used by many members of our society, including students, teachers, and researchers. Although we are aware of the sizeable environmental costs of AI development (Driessens & Pischetola, 2024) and the known existential risks of unregulated AI (OECD, 2024), pragmatically speaking, we do not believe that resisting the development of AI is a tenable solution. Burying our heads in the sand reduces our ability to realise the potential benefits of AI in learning and to prepare students for the critique of AI systems.

This new awareness in scholarship, policy, and practice reflects the idea that GenAI is not by design a threat to assessment (Pearce & Chiavaroli, 2023), and not all use of AI tools is automatically misconduct (Eaton, 2024). Although there are gaps in the publicly available university policies on the use of GenAI in writing (Moorhouse et al., 2023), there is evidence of moving to an approach that allows for the use of GenAI in writing (Perkins & Roe, 2023), scholarly publishing (Perkins & Roe, 2024a), and research (Perkins & Roe, 2024b).

Similarly, assessment redesign has become a major focus in preparing for a future AI-enabled world, as several authors have suggested (Mao et al., 2024; Thanh et al., 2023; Thompson et al., 2023; Xia et al., 2024). Single examinations or essays are increasingly being recognised as untenable for assessing learning (Gorichanaz, 2023). Among the topics raised in the literature on assessment redesign, common themes include focusing on student-centred (Hsiao et al., 2023), authentic (Rasul et al., 2024), learning instead of the end product, higher-order thinking skills (Smolansky et al., 2023), and evaluative judgement (Bearman et al., 2024). This recognition has helped advance frameworks for integrating AI into assessments such as the AIAS. However, the AIAS is not the only framework available for integrating AI into education, and multiple approaches have been proposed. In medical education for example, Pearce & Chiavaroli (2023) suggest that medical education rely on an 'assisted' and 'unassisted' dichotomy, one which allows external resources including GenAI and one which does not, with a refocusing on neglected assessment types such as oral examinations. Other useful frameworks include the two-lane approach (Liu & Bridgeman, 2023), six-lane approach (Steel, 2024), and traffic light approach (Cotterell, 2024).

## The Original AI Assessment Scale

Before discussing our updates, we will present the original AI Assessment Scale (Perkins et al., 2024). A core understanding of the original AIAS is that banning AI is not a productive or tenable approach. Lodge's (2023) assertion that the future of assessment requires students to be able to participate an in AI-reliant society sums up this position well: whether we personally use or refuse GenAI in our classrooms, students will encounter the technology in many aspects of their lives. At the same time, there is a very practical problem that causes significant distress for practitioners and colleagues around the world. In other words, how are we to support academic integrity and ensure that students' submitted work reflects their true abilities? Our answer was that the only way we could know for sure was through open dialogue with students and supporting clarity and transparency in the use of GenAI in assessment.

Our second challenge was that the AIAS had to be fit for purpose, flexible, and easy to understand and explain. Assessment varies significantly across disciplines, levels, and institutional norms, and we had to devise something that would not only suit our own subject areas (broadly social sciences), but also be adaptable to many assessment contexts. Third, we wanted to address the education sector as a whole, and given our backgrounds as both higher education and K-12 educators, we did not want to limit the AIAS to higher education but also make it applicable to K-12 education.

The primary purpose of the original AIAS was to facilitate meaningful dialogue between educators and students about the appropriate use of GenAI in assessments. Rather than adopting a simplistic binary approach of either allowing or prohibiting AI use, we recognised that a more nuanced framework was needed to help both educators and students understand the technology's potential strengths and limitations. Given that GenAI was an emerging technology, we structured the framework into five distinct levels of use, allowing educators and students to map these levels onto their existing understanding of assessment types, while providing clear guidance for implementation.

**Theoretical underpinning of the AIAS**

At the time of conceiving the framework, we discussed the theoretical underpinnings of our approach to using it in learning and assessment, but we did not fully expound this in our original paper. Broadly, our view of the AI Assessment Scale is grounded in social constructivist principles, particularly Vygotsky's (1978) understanding of learning as being mediated through social interaction. From this perspective, GenAI tools can be viewed as mediating technologies that might support students' knowledge construction, similar to how Vygotsky conceptualised language and cultural tools as mediators of higher mental functions. The different levels of the AIAS reflect varying degrees of mediated learning, from independent demonstrations of knowledge to full integration of AI as a collaborative partner in the learning process. Importantly, this perspective of AI use centres on the social experiences of the learner and the educator and avoids what we consider to be problematic understandings of AI chatbots as tutors for personalised learning where students learn in isolation from their peers (Bewersdorff et al., 2023).

Central to this theoretical framing is Vygotsky's (1978) concept of the Zone of Proximal Development (ZPD), the gap between what learners can achieve independently and what they can achieve with support. GenAI tools can function within this zone by providing scaffolding that helps bridge the gap between their current and potential performance. However, unlike traditional scaffolding, which is gradually removed over time, AI tools remain available as part of the learner's toolkit, requiring careful consideration of how they can support rather than replace learning. This aligns with the AIAS's emphasis on transparency and appropriate technology use, rather than restriction and control.

The social constructivist framework also helps explain why the AIAS emphasises supporting rather than restricting AI use. From this perspective, learning occurs through multiple interactions with educators and peers, technologies, and broader environments. Knowledge is actively constructed through these interactions rather than being passively received. The AIAS acknowledges this by creating spaces for students to engage critically with AI tools while developing their understanding rather than attempting to enforce arbitrary boundaries between AI-assisted and independent work.

The AIAS, as originally conceived, is presented in Table 1.

| | | |
|---|---|---|
| 1 | **NO AI** | The assessment is completed entirely without AI assistance. This level ensures that students rely solely on their knowledge, understanding, and skills.<br>**AI must not be used at any point during the assessment.** |
| 2 | **AI-ASSISTED IDEA GENERATION AND STRUCTURING** | AI can be used in the assessment for brainstorming, creating structures, and generating ideas for improving work.<br>**No AI content is allowed in the final submission.** |
| 3 | **AI-ASSISTED EDITING** | AI can be used to make improvements to the clarity or quality of student created work to improve the final output, but no new content can be created using AI.<br>**AI can be used, but your original work with no AI content must be provided in an appendix.** |
| 4 | **AI TASK COMPLETION, HUMAN EVALUATION** | AI is used to complete certain elements of the task, with students providing discussion or commentary on the AI-generated content. This level requires critical engagement with AI generated content and evaluating its output.<br>**You will use AI to complete specified tasks in your assessment. Any AI created content must be cited.** |
| 5 | **FULL AI** | AI should be used as a 'co-pilot' in order to meet the requirements of the assessment, allowing for a collaborative approach with AI and enhancing creativity.<br>**You may use AI throughout your assessment to support your own work and do not have to specify which content is AI generated.** |

*Table 1 The Original AI Assessment Scale*

**Critique of the AIAS**

The first iteration of the AIAS, as a flexible and useful tool to challenge an urgent threat, has been demonstrated to be effective in empirical studies for improving student outcomes, supporting thoughtful assessment redesign, and reducing misconduct through reframing of GenAI use (Furze et al., 2024). However, through practical use among multiple contexts, from K-12 to Higher Education, and across five continents, we have been able to develop a deep understanding of the areas in which the AIAS can be further developed considering the changing nature of GenAI.

One of the areas identified for development of the AIAS is the separation of the framework to differentiate between K-12 and higher education (Kılınç, 2024). While we recognise that the AIAS is broad in scope, the intention was to provide a tool that could form a basis for adaptation in multiple contexts at different educational levels. To this end, while it would be unlikely at a younger level to encourage the unscaffolded use of GenAI tools to meet assessment tasks, it is dependent on the educator to make such decisions and use the framework in line with their context and goals. Furthermore, adaptations of the AIAS can be developed freely in different contexts, such as in our own adaptation to the English for Academic Purposes (EAP) classroom (Roe, Perkins, & Tregubova, 2024). To clarify this point, we contend that the AIAS is a starting point for educators to consider how to reform their assessments, and is not a prescriptive tool.

A further criticism of the original AIAS was the use of traffic light colours, ranging from red through three shades of yellow/amber to green. It has been noted that this use of colours is problematic, with red potentially implying a 'stop' or ban approach, and green having a positive implication such as success (Kılınç, 2024). There are also potential connotations of red as "bad" and green as "good" which were unintended: we do not believe that any type of assessment is inherently better or worse than any other, only that they should vary dependent on context. In the updated AIAS, we decided to move away from this system and provide a more neutral palette of colours, first to avoid such implications, and second to ensure legibility and accessibility. We also note that there are legitimate defences for the use of a red-amber-green 'traffic light' system, such as for ease of clarity with younger students (Cotterell, 2024) and in situations where supervised and more prescriptive assessments may be possible, such as in the comparatively smaller class sizes and more often face to face settings of K-12 education versus the large and more commonly online or hybrid higher education cohort. As with all AIAS resources, we encourage educators to adapt and make their own decisions based on the context, expertise, and needs of the learners.

A further issue we have noted in the practical implementation of the AIAS is when it has been used for superficial changes to assessment – something discussed in detail towards the end of this article. Pratschke (2024) identified that the AIAS is beneficial for discussing GenAI use, but that eventually, full transformation of assessment will be required to cope with the changes posed by these technologies. At the time of development, the original AIAS was intended to be used to restructure assessment in depth; however, because of the urgent need to accommodate new technologies, it also gained traction as a way to augment existing practices. In refining the AIAS further, we have attempted to orient it toward a framework that provides a structured basis for a bottom-up transformational process.

A critique of scale-based approaches to AI integration comes from Liu and Bridgeman's (2023) two-lane approach to AI assessment developed at the University of Sydney. They proposed that assessments should be divided between secured 'lane 1' assessments that assure learning outcomes, and 'lane 2' assessments that embrace human-AI collaboration. They argue there is no "*viable middle ground*" between these lanes - any unsecured assessment must assume AI use will occur and cannot be controlled. While we acknowledge this reality in unsupervised environments, we contend that the AIAS serves a different purpose. Rather than attempting to control AI use, the AIAS supports transparent conversations between educators and students about appropriate AI use while simultaneously encouraging assessment redesign to account for the specific level of the scale selected. We argue that in these relatively early days of the technology, both educators and students could benefit from the clarity of different levels of the AIAS rather than the broad and potentially ambiguous dichotomy of the two lanes.

## Evolution of the AIAS:

**Key revisions and rationale**

Drawing from critiques, implementation experiences, and considerations of assessment validity and transparency, we made several significant refinements to the second iteration of the AIAS. Most visibly, we replaced the traffic light colour scheme and introduced an additional circular representation of the framework (shown in Figure 2). This circular design emphasises that no level is inherently better than others, aligning with our focus on appropriate rather than restrictive AI use based on specific learning outcomes and contexts. The framework was also updated to reflect the rapid evolution of GenAI technologies, particularly acknowledging the increasing prevalence of multimodal content generation across text, images, audio, and video.

The revisions reflect a deeper understanding of assessment validity and transparency in the GenAI era. Dawson et al. (2024) argue that validity in assessment must represent a student's true level of knowledge and skills, serving as a social imperative that integrates fairness and inclusivity. Following this principle, we removed the previous Level 3 requirement for students to include an appendix showing their original work, acknowledging that the current detection methods cannot reliably verify such submissions. This aligns with Bearman et al.'s (2024) emphasis on evaluative judgment, focusing on students' ability to assess and identify quality work, rather than attempting to control technology use. The original version of the AIAS was recognised by the Australian Tertiary Education Quality and Standards Agency (TEQSA) (Lodge, 2024) as a potential tool to help delineate the appropriate uses of AI in assessment tasks, and was commended in the Prof Tracey Bretag Prize for Academic Integrity (Studiosity, 2024) This early validation of the framework's value in promoting transparency informed our continued emphasis on clear communication in this revised version.

Text-generation capabilities are now recognised at lower levels of the scale, reflecting the reality that co-writing with AI has become a common practice and may support learning when appropriately scaffolded (Dhillon et al., 2024; Nguyen et al., 2024; Wiboolyasarin et al., 2024). At the highest level of the scale, we have strengthened the emphasis on co-design between educators and students, recognising that, as these technologies evolve, the most innovative applications often emerge from the collaborative exploration of AI tools in educational contexts. These modifications maintain the AIAS's core purpose as a framework for transparent dialogue between students and educators, while adapting to technological developments and emerging educational practices. The revised AIAS is shown in figure 1. Figure 2 shows the circular representation of the scale.

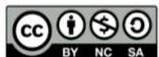

*Figure 1: The Revised AI Assessment Scale*

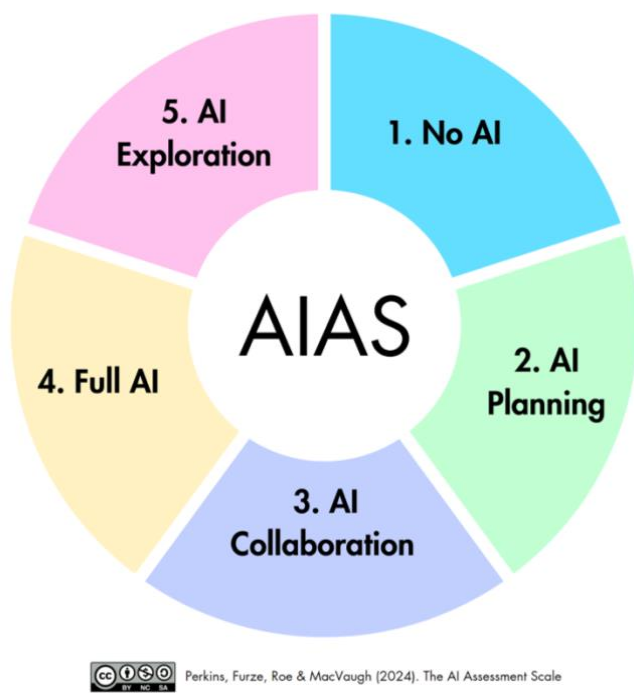

*Figure 2: Circular Representation of the Revised AI Assessment Scale*

To support the understanding of the various levels of AIAS, we provide examples of assessments which may be suitable at each level through a series of "vignettes" which are composites of the authors' experiences and feedback from educators around the world who have used the AIAS. Vignettes can provide real-life examples rather than mere suggestions and can help educators contextualise and frame others' experiences through their own expertise (Sampson & Johannessen, 2020).

**Level 1: No AI**

This level represents a traditional and controlled assessment environment in which AI tools are strictly prohibited and prohibition is enforced by removing access to digital technologies. To complete a task, students must rely entirely on their existing knowledge, skills, and understanding. This approach ensures that the assessment directly measures the students' unaided capabilities and core competencies. It is particularly suitable for evaluating foundational knowledge, critical thinking skills, and the ability to articulate ideas, without technological assistance.

The revision of Level 1 represents a significant shift from our original conceptualisation, moving beyond the simple prohibition of AI to a more nuanced understanding of when and why an AI-free assessment is appropriate. This refinement emerged from practical implementation experiences and feedback from educators implementing the AIAS, which revealed that merely declaring an AI-free assessment without environmental controls was increasingly untenable in contemporary educational settings. Therefore, a key modification in the revised framework is the explicit emphasis on controlled environments, reflecting Dawson et al.'s (2024) argument that assessment validity must take precedence over traditional notions of academic integrity. In practice, this means that Level 1 assessments should occur in supervised settings where the absence of AI can be assured, rather than relying on honour systems or detection tools for take-home work. This presents challenges for online and distance-learning environments, where truly AI-free conditions are difficult to guarantee without physical supervision. Some learning outcomes may be better served by allowing strategic AI use, thereby making other levels of the framework more appropriate.

A critical consideration in the Level 1 implementation is the balance between security and accessibility. While these assessments preclude AI use, they must accommodate students who require assistive technology. This requires a careful distinction between AI tools that can compromise assessment validity, and assistive technologies that enable equitable participation. The framework acknowledges that blanket bans of technology can create unnecessary barriers for students with disabilities and that assessment design should consider accessibility from the outset rather than as an afterthought. As we progress into an increasingly 'postplagiarism' world, (Eaton, 2023), the ability of educational institutions to truly create an entirely controlled assessment environment may soon disappear entirely. This requires major reconsideration of the relationship between assessment security and validity.

*Example of Level 1 Assessment*

In a writing course, a faculty member determines that a particular rhetorical skill cannot be validly assessed if students have access to GenAI. The students' demonstration of these skills, such as control of sentence structure, grammar, and rhetorical devices, would be obfuscated by software such as Grammarly, ChatGPT, or Microsoft Word with Copilot. As such, the faculty determines that a "No AI" assessment is required. The students complete a written exercise by hand, under supervision. For students requiring assistive technologies, word processor technologies (such as an e-ink tablet with a keyboard) or a scribe are provided.

**Level 2: AI-Assisted Planning**

The revision of Level 2 represents a shift in the recognition of how GenAI can support the planning and ideation phases of assessments. While the original framework emphasised basic brainstorming capabilities, the revised Level 2 acknowledges AI's broader role in supporting preliminary work while focusing on students' active development of ideas. Simply declaring that AI can be used 'only for planning' in traditional assessments is insufficient, and tasks must be explicitly designed to evaluate students' ability to develop and refine ideas generated through AI assistance. This will likely involve process-focused tasks, as opposed to output-focused tasks.

A key consideration in Level 2 implementation is how AI can support rather than restrict learning processes. Rather than attempting to control AI use through unreliable detection methods as we discuss earlier, Level 2 emphasises supporting students in developing their planning and ideation skills with AI as a collaborative tool. This requires carefully structured assessment components that distinguish between AI-supported planning phases and independent development while acknowledging that strict delineation between these phases may not be technically feasible. Assessment design should acknowledge that while we cannot technically restrict AI use to planning stages, we can create tasks that meaningfully engage students in demonstrating their development of ideas.

*Example of Level 2 Assessment*

In a Media or Film course, an interim assessment is developed in which students create a comprehensive storyboard and plan for a documentary to be produced as a final project. The assessment task is focused on the storyboard and contributes 25% of the total final grade for the unit, with the remaining 75% coming from the documentary production. For the storyboard, the course leader decides that AI use at Level 2 is acceptable because it is possible to make a valid judgement of the students' abilities to plan and design a shot sequence, whether they use AI or not. Furthermore, this task can be supported by a variety of multimodal GenAI tools, including research (e.g., Perplexity, ChatGPT), sketch-to-image generation (e.g., Firefly,

ControlNet) and image-to-video generation to create moving storyboards (e.g., Runway, Kling, Sora). Students are explicitly taught how to use these tools, and are assessed on their understanding of storyboarding conventions, shot sequencing, pace, and other elements. Faculty members may also decide to assess the use of AI in this process.

**Level 3: AI-Assisted Task Completion**

The revision of Level 3 represents a significant shift from our original conception, moving beyond simple editing, to acknowledge the reality of AI-assisted drafting and composition. While the original framework attempted to restrict AI use to surface-level improvements of the original work of students, the revised Level 3 recognises that attempting to enforce such limitations is neither practical nor pedagogically sound. Instead of focusing on authorship verification, Level 3 emphasises the development of critical evaluation skills and the maintenance of student voice in AI-assisted writing. Although technological solutions exist that purport to show how students develop their work, such as document tracking or version control systems, they are fallible and can be circumvented by determined students. More importantly, the emphasis on 'proving' original authorship can create inequitable conditions, particularly disadvantaging students without access to more advanced (and paid for) GenAI models, or technical knowledge of how to effectively integrate AI assistance while maintaining apparent independence.

Supporting students at this level must consider how to overcome the *illusion of finality* - the tendency to accept AI-generated text as complete and authoritative. Students must be supported in understanding the limitations of AI systems and recognising that their own knowledge, critical thinking skills, and subject expertise are essential for producing high-quality academic work. This involves developing critical AI literacy (CAIL) skills (Roe, Furze, & Perkins, 2024), identifying potential errors or biases, and effectively integrating AI-generated content with their own insights and understanding to ensure that their own voice is retained. The formative use of AI detection tools might also support this process, not as a means of catching misconduct but as a way of helping students understand how their writing style may be influenced by AI and how to retain their distinctive voice.

*Example of Level 3 Assessment*

In science faculty, educators determine that students could effectively use GenAI as part of the write-up of an earlier practical experiment. Prior to the assessment task, students collected data and conducted practical work. GenAI may be used in the write-up of these data in numerous ways. These include: the use of AI-assisted editing tools (e.g., Grammarly and Microsoft Copilot in MS Word); the use of GenAI for data analytics and assistance with understanding data (e.g., Microsoft Copilot in MS Excel, Claude or ChatGPT's data interpretation functions); and the synthesis of existing writing in the required style to produce a "guide" for students to follow in their own writing. The faculty member explicitly teaches students how to use these GenAI technologies in the way that they consider to be appropriate.

**Level 4: Full AI**

Level 4 represents a significant shift in how AI is integrated into assessments, focusing on the strategic deployment of AI tools to achieve specified learning outcomes. While this level previously represented the highest degree of AI integration in our original scale, its refinement acknowledges that as AI capabilities expand, particularly with multimodal tools, students need opportunities to develop a sophisticated understanding of when and how to effectively leverage these technologies. Level 4 supports students' ability to direct AI tools while demonstrating critical thinking and subject knowledge. Unlike lower levels, where AI use is more constrained, Level 4 assessments evaluate how effectively students leverage AI to solve problems and demonstrate understanding. This aligns with Dawson et al.'s (2024) emphasis on assessment validity–the focus shifts from controlling AI use to ensuring that students demonstrate genuine learning through their strategic deployment of AI tools.

A key consideration at this level is equity of access to AI tools. While providing specific tools through shared institutional accounts may help address some inequities, educators must carefully design assessments that remain valid, regardless of students' access to advanced AI models. This might involve providing institutional AI tools or designing tasks that can be completed effectively using freely available resources. Although this may help address equity concerns, we acknowledge that it remains an imperfect solution. Nevertheless, transparency about permitted or required tools helps to create clearer expectations while potentially reducing access barriers. Level 4 opens opportunities for multimodal assessment designs that reflect expanding GenAI capabilities. Beyond text generation, students might engage with AI-assisted video creation, synthetic media production, or combinations of multiple modes to achieve outcomes that would be greater than those realistically created by the student otherwise, given the time restrictions.

*Example of Level 4 Assessment*

The teachers of a third-year university computer science course determine that students would be best served by learning and having access to the most cutting-edge AI-assisted coding tools, many of which are already deployed in the software design industry (e.g., Github Copilot, Cursor, or OpenAI's o1 model). The institution provides students with access to these technologies during this course. Students may complete this assessment using any AI-assisted technologies available to them, but they are explicitly taught the use of the software that the institution provides access to. Students are assessed on both their abilities to design solutions to the problem in the task, and in their use of AI systems.

**Level 5: AI Exploration**

Level 5 represents a forward-looking vision of assessment design, centred not simply on incorporating AI tools, but on using them to reimagine what academic tasks can entail. Unlike earlier levels, which focus on controlling, channelling, or strategically deploying AI, this level encourages the co-creation and development of entirely new methods, media, and outputs that challenge conventional disciplinary boundaries. It builds upon the foundational understanding of AI use outlined in the lower levels of the framework, particularly the emphasis on authenticity, critical engagement, and strategic tool selection found in Level 4, but it extends these principles into future-oriented scenarios that may not yet be fully feasible with current technology.

At this stage, students are not only using AI tools to enhance their work; they are expected to conceptualise and implement new AI applications that go beyond existing templates and straightforward multimodal generation. This might mean creating world models or integrating advanced synthetic media production into an assessment task, which transcends the more direct and instrumental uses of AI at Level 4. While Level 4 acknowledges the arrival of advanced multimodal systems, Level 5 imagines a space where these tools become the raw materials from which students and educators collaboratively sculpt novel approaches to enquiry, whether it involves simulating complex social phenomena, generating bespoke datasets that do not currently exist, or building adaptive systems that respond dynamically to user input and changing conditions. The role of the educator also changes, moving from an authority that defines the parameters of the task to a collaborator that shares ownership of the learning process. This co-creation aspect, suggested in earlier treatments of AI integration but not fully realised, aligns with the need for flexibility, adaptability, and openness to new forms of assessment in a world where digital and human cognition intertwine.

The implementation of Level 5 assessments requires careful consideration of equity issues, as highlighted earlier in our Level 4 discussion. However, at this level, the focus shifts from providing equal access to tools to ensuring equal opportunities for innovation and experimentation. This might involve the institutional provision of advanced AI capabilities through self-hosted GenAI models or API access or the development of custom tools specific to disciplines or research areas. As Bearman et al. (2024) suggest, evaluative judgement becomes crucial here–students must not only use AI effectively, but also critically assess its contributions to their innovative work.

This level may be most relevant to advanced undergraduate projects, postgraduate coursework, doctoral research, or cutting-edge independent endeavours at the secondary level, such as the extended projects of the International Baccalaureate. It challenges the assumption that AI must be contained or limited. Instead, it posits that the measure of learning lies in how well students can harness, adapt, and extend AI capabilities for new intellectual, aesthetic, or practical ends. Such assessments might respond to the evolving state of AI, acknowledging that as brain-computer interfaces (BCIs) and other technologies develop (Eaton, 2023), traditional assessment controls or even conventional understanding of "authentic" tasks may rapidly become obsolete. In this way, Level 5 is deliberately futuristic, anticipating that new toolkits, disciplinary challenges, and ways of knowing will emerge and that assessment must evolve to ensure students are equipped to navigate this uncertain terrain.

*Example of Level 5 Assessment*

In a creative arts specialist institution, educators and students are exploring the limits of multimodal generative AI, including image recognition and segmentation, to understand how AI can be used to support contemporary dance. Some students work on a system that can provide real-time feedback on performers' movements, posture, and position, whereas others explore the use of machine learning to create procedurally generated lighting and effects which map to a dancer's movement. The educator invites external speakers who work in both dance and GenAI to help the students refine their ideas.

## Conclusion

The revised version of the AIAS represents a significant evolution in our understanding of how to integrate GenAI into educational assessments. Key changes include moving away from the traffic light colour scheme to avoid implying hierarchical values, recognising text generation capabilities at earlier levels of the scale considering evidence regarding both AI detection limitations and potential learning benefits, and adjusting Level 5 to acknowledge the rapid advancement of these technologies. The framework now better reflects the complex reality of assessment in an AI-enabled world while maintaining its core purpose of supporting both transparent dialogue between educators and students and the effective redesign of assessments.

However, we must acknowledge that frameworks for AI integration in education cannot ignore the broader ethical and environmental impacts of these technologies. The planetary costs of training and running large language models are substantial (Driessens & Pischetola, 2024), and concerns regarding human rights and values in AI development remain pressing (Eaton, 2024; Rudolph et al., 2024). The challenge of maintaining transparency in AI use is highlighted by research showing students' reluctance to declare AI use even when permitted (Gonsalves, 2024), suggesting that deeper cultural shifts may be needed in how we view and discuss AI assistance in academic work.

While early evidence suggests that the AIAS can help improve student outcomes, reduce academic misconduct and support educators in effective assessment design (Furze et al., 2024), further empirical validation is crucial. Future research should examine both the framework's effectiveness across different educational contexts and the broader impact of GenAI integration in education (Koh & Doroudi, 2023). This research should include qualitative investigations of student and educator experiences with the framework, as well as quantitative measures of its impact on learning outcomes and assessment validity. The addition of Level 5 represents our attempt to future-proof the framework, acknowledging Dawson and Bearman's (2020) observation that forward-looking assessment requires some prediction of future developments. While we cannot perfectly anticipate technological advancement, Level 5 provides flexibility for incorporating emerging AI capabilities while maintaining a focus on human creativity and critical engagement.

We emphasise that the AIAS is not prescriptive; institutions should adapt it to their specific contexts and needs. However, the reality that GenAI is now an established part of the educational landscape cannot be ignored. Rather than attempting to restrict or control AI use through increasingly ineffective technical measures, educators must engage in open dialogue with students about appropriate AI integration in assessments. The AIAS provides a structured way for these conversations to occur, while acknowledging both the opportunities and challenges of GenAI in education. Ultimately, the success of any framework for AI integration will depend not only on its technical robustness or theoretical grounding but also on how effectively it supports meaningful learning in an increasingly AI-enabled world. As we continue to gather evidence and refine our approaches, maintaining a focus on assessment validity, the impact of student learning, and ethical considerations are crucial. The AIAS represents one step toward this goal, but much work remains in understanding how best to prepare students for a future in which human and artificial intelligence are increasingly intertwined.

**Acknowledgements**

We are grateful for the support and guidance provided by Jason MacVaugh during the redevelopment of the revised version of the AIAS. We thank all educators worldwide who have provided comments and critiques on the two versions of the scale, both formally and informally.

**AI usage disclaimer**

GenAI tools were used for text creation, revision, and editorial purposes throughout the production of the manuscript. Tools used were ChatGPT (GPT-4o) and Claude 3.5 Sonnet. The authors reviewed, edited, and take responsibility for all outputs of the tools used.

# Bibliography

Bearman, M., Tai, J., Dawson, P., Boud, D., & Ajjawi, R. (2024). Developing evaluative judgement for a time of generative artificial intelligence. *Assessment & Evaluation in Higher Education*, *0*(0), 1–13. https://doi.org/10.1080/02602938.2024.2335321

Bewersdorff, A., Zhai, X., Roberts, J., & Nerdel, C. (2023). Myths, mis- and preconceptions of artificial intelligence: A review of the literature. *Computers and Education: Artificial Intelligence*, *4*, 100143. https://doi.org/10.1016/j.caeai.2023.100143

Cotterell, A. (2024, September 3). A defence of the traffic light metaphor. *Adrian Cotterell*. https://adriancotterell.com/2024/09/04/a-defence-of-the-traffic-light-metaphor/

Cotton, D. R. E., Cotton, P. A., & Shipway, J. R. (2023). Chatting and cheating: Ensuring academic integrity in the era of ChatGPT. *Innovations in Education and Teaching International*, *0*(0), 1–12. https://doi.org/10.1080/14703297.2023.2190148

Crawford, J., Cowling, M., & Allen, K.-A. (2023). Leadership is needed for ethical ChatGPT: Character, assessment, and learning using artificial intelligence (AI). *Journal of University Teaching & Learning Practice*, *20*(3), 02.

Dawson, P., & Bearman, M. (2020). Concluding Comments: Reimagining University Assessment in a Digital World. In M. Bearman, P. Dawson, R. Ajjawi, J. Tai, & D. Boud (Eds.), *Re-imagining University Assessment in a Digital World* (pp. 291–296). Springer International Publishing. https://doi.org/10.1007/978-3-030-41956-1_20

Dawson, P., Bearman, M., Dollinger, M., & Boud, D. (2024). Validity matters more than cheating. *Assessment & Evaluation in Higher Education*, *49*(7), 1005–1016. https://doi.org/10.1080/02602938.2024.2386662

Dhillon, P. S., Molaei, S., Li, J., Golub, M., Zheng, S., & Robert, L. P. (2024). Shaping Human-AI Collaboration: Varied Scaffolding Levels in Co-writing with Language Models. *Proceedings of the 2024 CHI Conference on Human Factors in Computing Systems*, 1–18. https://doi.org/10.1145/3613904.3642134

Driessens, O., & Pischetola, M. (2024). Danish university policies on generative AI: Problems, assumptions and sustainability blind spots. *MedieKultur: Journal of Media and Communication Research*, *40*(76), Article 76. https://doi.org/10.7146/mk.v40i76.143595

Eaton, S. E. (2023). Postplagiarism: Transdisciplinary ethics and integrity in the age of artificial intelligence and neurotechnology. *International Journal for Educational Integrity*, *19*(1), 23. https://doi.org/10.1007/s40979-023-00144-1

Eaton, S. E. (2024). Future-proofing integrity in the age of artificial intelligence and neurotechnology: Prioritizing human rights, dignity, and equity. *International Journal for Educational Integrity*, *20*(1), Article 1. https://doi.org/10.1007/s40979-024-00175-2

Elkhatat, A. M., Elsaid, K., & Almeer, S. (2023). Evaluating the efficacy of AI content detection tools in differentiating between human and AI-generated text. *International Journal for Educational Integrity*, *19*(1), 17. https://doi.org/10.1007/s40979-023-00140-5

Farazouli, A., Cerratto-Pargman, T., Bolander-Laksov, K., & McGrath, C. (2023). Hello GPT! Goodbye home examination? An exploratory study of AI chatbots impact on university teachers' assessment practices. *Assessment & Evaluation in Higher Education*, *0*(0), 1–13. https://doi.org/10.1080/02602938.2023.2241676

Furze, L. (2024, December 9). *AI Assessment Scale (AIAS) Translations from Around the World*. Leon Furze. https://leonfurze.com/2024/12/09/ai-assessment-scale-aias-translations-from-around-the-world/

Furze, L., Perkins, M., Roe, J., & MacVaugh, J. (2024). The AI Assessment Scale (AIAS) in action: A pilot implementation of GenAI-supported assessment. *Australasian Journal of Educational Technology*. https://doi.org/10.14742/ajet.9434

Gonsalves, C. (2024). Addressing student non-compliance in AI use declarations: Implications for academic integrity and assessment in higher education. *Assessment & Evaluation in Higher Education*, *0*(0), 1–15. https://doi.org/10.1080/02602938.2024.2415654

Gorichanaz, T. (2023). Accused: How students respond to allegations of using ChatGPT on assessments. *Learning: Research and Practice*, *9*(2), 183–196. https://doi.org/10.1080/23735082.2023.2254787

Head, A., & Willis, S. (2024). Assessing law students in a GenAI world to create knowledgeable future lawyers. *International Journal of the Legal Profession*, *0*(0), 1–18. https://doi.org/10.1080/09695958.2024.2379785

Hsiao, Y.-P., Klijn, N., & Chiu, M.-S. (2023). Developing a framework to re-design writing assignment assessment for the era of Large Language Models. *Learning: Research and Practice*, *9*(2), 148–158. https://doi.org/10.1080/23735082.2023.2257234

Kılınç, S. (2024). *Comprehensive AI Assessment Framework: Enhancing Educational Evaluation with Ethical AI Integration* (arXiv:2407.16887). arXiv. http://arxiv.org/abs/2407.16887

Koh, E., & Doroudi, S. (2023). Learning, teaching, and assessment with generative artificial intelligence: Towards a plateau of productivity. *Learning: Research and Practice*, *9*(2), 109–116. https://doi.org/10.1080/23735082.2023.2264086

Liang, W., Yuksekgonul, M., Mao, Y., Wu, E., & Zou, J. (2023). GPT detectors are biased against non-native English writers. *arXiv Preprint arXiv:2304.02819*.

Liu, D., & Bridgeman, A. (2023, July 12). What to do about assessments if we can't out-design or out-run AI? *What to Do about Assessments If We Can't out-Design or out-Run AI?* https://educational-innovation.sydney.edu.au/teaching@sydney/what-to-do-about-assessments-if-we-cant-out-design-or-out-run-ai/

Lodge, J., Howard, S., Bearman, M., & Dawson, P. (2023). *Assessment reform for the age of Artificial Intelligence*. Tertiary Education Quality and Standards Agency.

Luo (Jess), J. (2024a). A critical review of GenAI policies in higher education assessment: A call to reconsider the "originality" of students' work. *Assessment & Evaluation in Higher Education*, *0*(0), 1–14. https://doi.org/10.1080/02602938.2024.2309963

Luo (Jess), J. (2024b). How does GenAI affect trust in teacher-student relationships? Insights from students' assessment experiences. *Teaching in Higher Education*, *0*(0), 1–16. https://doi.org/10.1080/13562517.2024.2341005

Mao, J., Chen, B., & Liu, J. C. (2024). Generative Artificial Intelligence in Education and Its Implications for Assessment. *TechTrends*, *68*(1), 58–66. https://doi.org/10.1007/s11528-023-00911-4

Moorhouse, B. L., Yeo, M. A., & Wan, Y. (2023). Generative AI tools and assessment: Guidelines of the world's top-ranking universities. *Computers and Education Open*, *5*, 100151. https://doi.org/10.1016/j.caeo.2023.100151

Newton, P. M., Summers, C. J., Zaheer, U., Xiromeriti, M., Stokes, J. R., Bhangu, J. S., Roome, E. G., Roberts-Phillips, A., Mazaheri-Asadi, D., Jones, C. D., Hughes, S., Gilbert, D., Jones, E., Essex, K., Ellis, E. C., Davey, R., Cox, A. A., & Bassett, J. A. (2024). *Can ChatGPT-4o really pass medical science exams? A pragmatic analysis using novel questions* (p. 2024.06.29.24309595). medRxiv. https://doi.org/10.1101/2024.06.29.24309595

Nguyen, A., Hong, Y., Dang, B., & Huang, X. (2024). Human-AI collaboration patterns in AI-assisted academic writing. *Studies in Higher Education*, *49*(5), 847–864. https://doi.org/10.1080/03075079.2024.2323593

OECD. (2024). *AI risks and incidents*. Artificial Intelligence Promises Tremendous Benefits but Also Carries Real Risks. Some of These Risks Are Already Materialising into Harms to People and Societies: Bias and Discrimination, Polarisation of Opinions, Privacy Infringements, and Security and Safety Issues. Trustworthy AI Calls for Governments Worldwide to Develop Interoperable Risk-Based Approaches to AI Governance and a Rigorous Understanding of AI Incidents and Hazards. https://www.oecd.org/en/topics/ai-risks-and-incidents.html

OpenAI. (2023, September 25). *GPT-4V(ision) system card*. GPT-4V(Ision) System Card. https://openai.com/index/gpt-4v-system-card/

OpenAI. (2024). *Introducing OpenAI o1 | OpenAI*. https://openai.com/index/introducing-openai-o1-preview/

Pearce, J., & Chiavaroli, N. (2023). Rethinking assessment in response to generative artificial intelligence. *Medical Education*, *57*(10), 889–891. https://doi.org/10.1111/medu.15092

Perkins, M. (2023). Academic Integrity considerations of AI Large Language Models in the post-pandemic era: ChatGPT and beyond. *Journal of University Teaching & Learning Practice*, *20*(2). https://doi.org/10.53761/1.20.02.07

Perkins, M., Furze, L., Roe, J., & MacVaugh, J. (2024). The Artificial Intelligence Assessment Scale (AIAS): A Framework for Ethical Integration of Generative AI in Educational Assessment. *Journal of University Teaching and Learning Practice*, *21*(06), Article 06. https://doi.org/10.53761/q3azde36

Perkins, M., & Roe, J. (2023). Decoding Academic Integrity Policies: A Corpus Linguistics Investigation of AI and Other Technological Threats. *Higher Education Policy*. https://doi.org/10.1057/s41307-023-00323-2

Perkins, M., & Roe, J. (2024a). Academic publisher guidelines on AI usage: A ChatGPT supported thematic analysis [version 2; peer review: 3 approved, 1 approved with reservations]. In *F1000Research* (Vol. 12, Issue 1398). https://doi.org/10.12688/f1000research.142411.2

Perkins, M., & Roe, J. (2024b). *Generative AI Tools in Academic Research: Applications and Implications for Qualitative and Quantitative Research Methodologies* (arXiv:2408.06872). arXiv. https://doi.org/10.48550/arXiv.2408.06872

Perkins, M., Roe, J., Postma, D., McGaughran, J., & Hickerson, D. (2023). Detection of GPT-4 Generated Text in Higher Education: Combining Academic Judgement and Software to Identify Generative AI Tool Misuse. *Journal of Academic Ethics*. https://doi.org/10.1007/s10805-023-09492-6

Perkins, M., Roe, J., Vu, B. H., Postma, D., Hickerson, D., McGaughran, J., & Khuat, H. Q. (2024). Simple techniques to bypass GenAI text detectors: Implications for inclusive education. *International Journal of Educational Technology in Higher Education*, *21*(1), 53. https://doi.org/10.1186/s41239-024-00487-w

Pratschke, B. M. (2024). Assessing Learning. In B. M. Pratschke (Ed.), *Generative AI and Education: Digital Pedagogies, Teaching Innovation and Learning Design* (pp. 91–108). Springer Nature Switzerland. https://doi.org/10.1007/978-3-031-67991-9_6

Prentice, F. M., & Kinden, C. E. (2018). Paraphrasing tools, language translation tools and plagiarism: An exploratory study. *International Journal for Educational Integrity*, *14*(1), 11. https://doi.org/10.1007/s40979-018-0036-7

Rasul, T., Nair, S., Kalendra, D., Balaji, M. S., Santini, F. de O., Ladeira, W. J., Rather, R. A., Yasin, N., Rodriguez, R. V., Kokkalis, P., Murad, M. W., & Hossain, M. U. (2024). Enhancing academic integrity among students in GenAI Era:A holistic framework. *The International Journal of Management Education*, *22*(3), 101041. https://doi.org/10.1016/j.ijme.2024.101041

Roe, J., Furze, L., & Perkins, M. (2024). *Funhouse Mirror or Echo Chamber? A Methodological Approach to Teaching Critical AI Literacy Through Metaphors* (arXiv:2411.14730). arXiv. https://doi.org/10.48550/arXiv.2411.14730

Roe, J., & Perkins, M. (2022). What are Automated Paraphrasing Tools and how do we address them? A review of a growing threat to academic integrity. *International Journal for Educational Integrity*, *18*(1), Article 1. https://doi.org/10.1007/s40979-022-00109-w

Roe, J., Perkins, M., & Ruelle, D. (2024). Is GenAI the Future of Feedback? Understanding Student and Staff Perspectives on AI in Assessment. *Intelligent Technologies in Education*. https://doi.org/10.70770/rzzz6y35

Roe, J., Perkins, M., & Tregubova, Y. (2024). *The EAP-AIAS: Adapting the AI Assessment Scale for English for Academic Purposes* (arXiv:2408.01075). arXiv. https://doi.org/10.48550/arXiv.2408.01075

Roe, J., Renandya, W. A., & Jacobs, G. M. (2023). A Review of AI-Powered Writing Tools and Their Implications for Academic Integrity in the Language Classroom. *Journal of English and Applied Linguistics*, *2*(1), 3.

Rudolph, J., Ismail, F. M. M., & Popenici, S. (2024). Higher Education's Generative Artificial Intelligence Paradox: The Meaning of Chatbot Mania. *Journal of University Teaching and Learning Practice*, *21*(06), Article 06. https://doi.org/10.53761/54fs5e77

Sadasivan, V. S., Kumar, A., Balasubramanian, S., Wang, W., & Feizi, S. (2023). *Can AI-Generated Text be Reliably Detected?* (arXiv:2303.11156). arXiv. https://doi.org/10.48550/arXiv.2303.11156

Shahriar, S., Lund, B. D., Mannuru, N. R., Arshad, M. A., Hayawi, K., Bevara, R. V. K., Mannuru, A., & Batool, L. (2024). Putting GPT-4o to the Sword: A Comprehensive Evaluation of Language, Vision, Speech, and Multimodal Proficiency. *Applied Sciences*, *14*(17), Article 17. https://doi.org/10.3390/app14177782

The AI Assessment Scale Revisited: A Framework for Educational Assessment A PREPRINT

Smolansky, A., Cram, A., Raduescu, C., Zeivots, S., Huber, E., & Kizilcec, R. F. (2023). Educator and Student Perspectives on the Impact of Generative AI on Assessments in Higher Education. *Proceedings of the Tenth ACM Conference on Learning @ Scale*, 378–382. https://doi.org/10.1145/3573051.3596191

Steel, A. (2024). *2 lanes or 6 lanes? It depends on what you are driving: Use of AI in Assessment*. https://www.education.unsw.edu.au/news-events/news/two-six-lanes-ai-assessment

Studiosity. (2024). *The Prof Tracey Bretag Prize for Academic Integrity*. https://www.studiosity.com/traceybretagprize

Thanh, B. N., Vo, D. T. H., Nhat, M. N., Pham, T. T. T., Trung, H. T., & Xuan, S. H. (2023). Race with the machines: Assessing the capability of generative AI in solving authentic assessments. *Australasian Journal of Educational Technology*, *39*(5), Article 5. https://doi.org/10.14742/ajet.8902

Thompson, K., Corrin, L., & Lodge, J. M. (2023). AI in tertiary education: Progress on research and practice. *Australasian Journal of Educational Technology*, *39*(5), Article 5. https://doi.org/10.14742/ajet.9251

Vygotsky, L. S. (1978). *Mind in Society: Development of Higher Psychological Processes*. Harvard University Press. https://doi.org/10.2307/j.ctvjf9vz4

Weber-Wulff, D., Anohina-Naumeca, A., Bjelobaba, S., Foltýnek, T., Guerrero-Dib, J., Popoola, O., Šigut, P., & Waddington, L. (2023). *Testing of Detection Tools for AI-Generated Text* (arXiv:2306.15666). arXiv. https://doi.org/10.48550/arXiv.2306.15666

Wiboolyasarin, W., Wiboolyasarin, K., Suwanwihok, K., Jinowat, N., & Muenjanchoey, R. (2024). Synergizing collaborative writing and AI feedback: An investigation into enhancing L2 writing proficiency in wiki-based environments. *Computers and Education: Artificial Intelligence*, *6*, 100228. https://doi.org/10.1016/j.caeai.2024.100228

Xia, Q., Weng, X., Ouyang, F., Lin, T. J., & Chiu, T. K. F. (2024). A scoping review on how generative artificial intelligence transforms assessment in higher education. *International Journal of Educational Technology in Higher Education*, *21*(1), 40. https://doi.org/10.1186/s41239-024-00468-z